# Bridging physiological and evolutionary time scales in a gene regulatory network

**Authors:**

Gwenaëlle Marchand[1,2], Vân Anh Huynh-Thu[3], Nolan C. Kane[4], Sandrine Arribat[5], Didier Varès[1,2], David Rengel[1,2], Sandrine Balzergue[5], Loren H. Rieseberg[6,7], Patrick Vincourt[1,2], Pierre Geurts[3], Matthieu Vignes[8] and Nicolas B. Langlade[1,2]

Corresponding author e-mail: nicolas.langlade@toulouse.inra.fr; phone: (+33) 5 61 28 54 58)

**Addresses:**

1: INRA, Laboratoire des Interactions Plantes-Microorganismes (LIPM), UMR441

2: CNRS, Laboratoire des Interactions Plantes-Microorganismes (LIPM), UMR2594, F-31326 Castanet-Tolosan, France

3: Department of Electrical Engineering and Computer Science and GIGA-R, Systems and Modeling, University of Liège, Liège, Belgium

4: Department of Ecology and Evolutionary Biology, University of Colorado at Boulder, Boulder, CO 80309, USA

5: INRA, Unité de Recherche en Génomique Végétale (URGV), UMR1165 – Université d'Evry Val d'Essonne – ERL CNRS 8196, CP 5708, F-91057 Evry Cedex, France

6: Department of Botany, University of British Columbia, Vancouver, British Colombia V6T 1Z4, Canada

7: Department of Biology, Indiana University, Bloomington, IN 47405, USA

8: INRA, Mathématiques et Informatique Appliquées (MIA), UPR875, F-31326 Castanet-Tolosan, France

**Summary**

Gene regulatory networks (GRN) govern phenotypic adaptations and reflect the trade-offs between physiological responses and evolutionary adaptation that act at different time scales. To identify patterns of molecular function and genetic diversity in GRNs, we studied the drought response of the common sunflower, *Helianthus annuus*, and how the underlying GRN is related to its evolution.

We examined the responses of 32,423 expressed sequences to drought and to abscisic acid and selected 145 co-expressed transcripts. We characterized their regulatory relationships in nine kinetic studies based on different hormones. From this, we inferred a GRN by meta-analyses of a Gaussian graphical model and a random forest algorithm and studied the genetic differentiation among populations ($F_{ST}$) at nodes.

We identified two main hubs in the network that transport nitrate in guard cells. This suggests that nitrate transport is a critical aspect of sunflower physiological response to drought. We observed that differentiation of the network genes in elite sunflower cultivars is correlated with their position and connectivity.

This systems biology approach combined molecular data at different time scales and identified important physiological processes. At the evolutionary level, we propose that network topology could influence responses to human selection and possibly adaptation to dry environments.

## Introduction:

Phenotype is shaped during an organism's life by its physiological and developmental responses to environmental conditions and across generations through evolutionary genetic adjustments to new environments. On the time scale of individual organisms, the phenotype can change rapidly due to gene regulatory networks (GRNs), which translate environmental and internal signals into physiological and developmental modifications. On an evolutionary time scale, such phenotypic modifications are based on changes in the genes composing the network that may alter this network at the structural or functional level.

Relating phenotypic modifications occurring at physiological and evolutionary time scales has been a major focus of evolutionary biologists for more than a century ((Osborn, 1896) and (Waddington, 1942) as well as more recently (Queitsch *et al.*, 2002); (Milo *et al.*, 2007)). Researchers have theorized (and later demonstrated) that physiological adaptation (for example, via regulation of gene expression or biochemical characteristics) can be replaced by an evolutionary change that becomes constitutive and alleviates the fitness costs associated with plasticity. This paradigm can be revisited in the context of a gene network. While gene regulatory networks are products of evolution, similar to other biological objects, GRNs also shape and constrain the evolvability of phenotypic responses to the environment.

Systems biology approaches, such as GRN inference, provide a global view of the different pathways that respond to environmental variation. A GRN is a genetic network based on gene expression levels (Wilkins, 2005). It describes transcriptional interactions and dynamics in response to environmental stressors, and therefore the GRN is key to understanding how organisms such as plants adapt to their environment.

Responses to environmental signals are often mediated through hormones. For example, in plants, abscisic acid (ABA) is produced during water stress in the vasculature and in the guard cells of the vegetative part of the plant (Boursiac *et al.*, 2013). Accordingly, the application of ABA induces the expression of genes involved in the response to dehydration and mimics drought stress. This interpretation has been confirmed by promoter analyses, which have demonstrated that these pathways share many targets (Shinozaki & YamaguchiShinozaki, 1997). The signals of different hormones interact and are integrated to convey environmental signals through the plant (Wilkinson *et al.*, 2012), suggesting that hormones should share transcriptomic targets.

Drought stress is a major abiotic factor that drives dramatic phenotypic changes in plants, including *Helianthus*, in which drought stress appears to constrain the colonization of new environments in the arid regions of the southwestern USA (Seiler & Rieseberg, 1997).

Therefore, the drought-stress GRN represents a tool for studying the interactions between organismal acclimation on the physiological time scale and population adaptation on the evolutionary time scale.

Several hormones mediate drought-stress responses; thus, the utilization of multiple hormonal treatments can elucidate the underlying GRN and highlight possible relationships between the genes involved. However, there are practical difficulties associated with the study of genetic networks. For example, the GRN identified could be biased toward interactions that have been previously detected in model species (Wilkins, 2005). To date, systems biology approaches, such as GRN inference, have been mostly restricted to model species, such as yeast (Dikicioglu *et al.*, 2011), *Drosophila* (Crombach *et al.*, 2012), or *Arabidopsis* (Ma *et al.*, 2007), and are typically performed under laboratory conditions. However, modeling dynamic biological processes requires time-series gene expression data that are relevant to both the biological process of interest and to the species targeted by the study. To understand genome function and evolutionary processes in an organism such as the sunflower, it is important to infer the GRN for the gene sets that are actually involved in the responses to a given environmental stress and to avoid the pitfall of using non-adapted model species data.

In this study, we used inference methods on sunflower data complemented with knowledge from Arabidopsis. These methods were specifically designed for time-series gene expression data and allowed us to reconstruct a sunflower GRN. The inferred GRN provides us a global view of the main physiological functions involved in the drought-stress responses occurring in the leaf, as well as their chronology.

On the evolutionary time scale, studying the underlying GRN for responses to environmental stresses such as drought can help explain how plants evolved to become better suited to their environments. Knowledge of gene's position in the GRN and its topological characteristics provides useful information about likely evolutionary constraints. For example, a highly connected gene is likely to be subject to many trade-offs, which would limit the accumulation of genetic diversity. Here, we identify correlations between network topology and genetic divergence between elite lines and landraces of sunflower and propose a mechanism to explain how sunflower genetic differentiation could be constrained in response to selective forces.

**Materials and Methods:**

Plant Material and growth conditions

Transcriptome interactions and dynamics were studied using the sunflower (*Helianthus annuus*) genotype XRQ. Plantlets were grown under hydroponic conditions in the previously described growth medium (Neumann *et al.*, 2000) in a growth chamber. After 14 days, the plantlets were treated by adding either mock solution (DMSO only in controls) or one of the following hormonal solutions : auxine (IAA); ethylene (ACC), gibberellic acid (GA3), salicylic acid (SA), methyl-jasmonate (MeJA), kinetin, ABA strigolactone (Stri) or Brassinol (Bras) Details about hormonal solutions are provided in Supporting Methods. First pairs of leaves was harvested at 0 (just before treatment), 1, 3, 6, 9, 24, and 48 hours after treatment, immediately frozen in liquid nitrogen, and stored at -80°C. The whole procedure was repeated three times for ACC, Bras, GA3, IAA, kinetin, SA, and Stri and four times for ABA and MeJA.

Gene selection

To identify genes which likely play a role in the drought GRN, a global transcriptomic approach was employed using an Affymetrix chip containing 32,423 probesets corresponding to sequences expressed in *Helianthus annuus* (Rengel *et al.*, 2012). Three different global transcriptomic datasets were analyzed and used to select genes. We selected genes that responded to at least two of the following conditions: (1) drought stress under field conditions; (2) drought stress under greenhouse conditions; and (3) 10 µM ABA application under hydroponic conditions.

The microarray data and analyses of the field and greenhouse conditions were previously reported by Rengel *et al.*. (2012). Under field conditions, plants of the Melody genotype were harvested at the post-flowering stage at a stress intensity level of 0.63 and 0.22 (ratio between evapotranspiration and maximal evapotranspiration) for irrigated and non-irrigated plants, respectively. Under greenhouse conditions, we recorded data from Melody pre-flowering plants at a fraction of transpirable soil water (FTSW) of 0.83 and 0.03 for the irrigated and non-irrigated plants, respectively.

The global transcriptomic data for the application of 10 µM ABA are new results and were obtained using the 6-hour treatment with ABA in the hydroponic experiment on the genotype XRQ (CATdb: AFFY_ABA_Sunflower or GEO accession: GSE22519). RNA quality verification, cDNA synthesis, and chip hybridization and washing were all performed using the Affymetrix platform at the INRA-URGV in Evry, France, following the protocol described in Rengel *et al.*, (2012). To identify the sunflower transcripts that were differentially regulated by ABA under our hydroponic conditions, the Affymetrix data were treated as previously described in Bazin *et al.,* (2011).

This list was extended to 181 genes with genes known to respond to the application of ABA or other hormones (literature (Boudsocq & Lauriere, 2005; Kawaguchi *et al.*, 2004; Miller *et al.*, 2009; Seki *et al.*, 2007; Umezawa *et al.*, 2010; Shinozaki & Yamaguchi-Shinozaki, 2007; Wang *et al.*, 2003; Wasilewska *et al.*, 2008; Rook *et al.*, 2006; Sirichandra *et al.*, 2009; Pastori & Foyer, 2002; Hirayama & Shinozaki, 2010; Li *et al.*, 2006; Bray, 2004; Valliyodan & Nguyen, 2006) or GO analysis).

Molecular analysis

The extraction of total RNA and cDNA synthesis were performed as described in Rengel *et al.* (2012). The expression levels of the 181 selected genes were analyzed in all samples by q-RT-PCR using the BioMark system (Fluidigm Corporation, San Francisco, CA, USA) as previously described (Spurgeon *et al.*, 2008). The q-RT-PCR results were analyzed following the $2^{ddCt}$ method (Livak & Schmittgen, 2001). Gene expression levels were normalized to the mean of previously validated reference genes (Rengel *et al.*, 2012) and to the corresponding control sample with the mock treatment. Detailed description of expression levels calculation is provided in the Supporting Methods.

Genetic differentiation among populations

Genetic polymorphisms of drought GRN genes were characterized in five different *Helianthus* populations, as described in a previous study (Renaut *et al.*, 2013): *H. argophyllus* (N=28), *H. petiolaris* (N=25), *H annuus* elite lines (N=9), *H. annuus* landrace lines (N=11), and wild *H. annuus* (N=11). Briefly, transcript sequences were obtained from young leave tissues with two RNAseq technologies (Roche 454 FLX and GAII Illumina pair-end sequencing 2x 100 bp). The transcript sequences were then aligned to the reference transcriptome using the Burros Wheeler Aligner (Li & Durbin, 2009). SNPs were called using the program SAMtools (Li *et al.*, 2009) with a minimum with Phred scaled genotype likelihoods of 30, corresponding to a genotyping accuracy of at least 99.9%. The population genetics statistic $F_{ST}$ was calculated between these populations for 89 of the 181 candidate genes using the R package HIERSTAT (Goudet, 2005). $F_{ST}$ is a widely used measure of genetic differentiation among populations.

GRN reconstruction

Missing values of gene expression (expressed as ΔΔCt) at time t=0 were imputed as values of 1. Other missing values (less than 1% of the values) were imputed with the R package IMPUTE by 10-nearest neighboring genes (Troyanskaya *et al.*, 2001).

After log transformation of the data, we performed an arithmetic mean over replicates to obtain a robust ΔΔCt expression value for each gene under each condition (time x treatment). We obtained nine datasets corresponding to the nine hormonal treatments and containing expression values for 145 genes with robust expression data at 7 different time points. From these nine datasets, we inferred 10 GRNs: one GRN from each hormonal treatment and a global GRN taking into account all treatments. Two complementary inference methods were used to achieve GRN predictions. The first method represents an extension of GENIE3 (Huynh-Thu *et al.*, 2010) and was based on the random forest method (RF, (Breiman, 2001)). In summary, each gene expression at time t+1 was successively considered as a target, and the method sought regulators of that gene via their expression at time t. Several regulator inclusion steps were successively performed: according to a variance reduction criterion in a regression tree framework, each step resulted in the inclusion in the model of the best regulator. The process was repeated on a randomized ensemble of trees, which made up the so-called random forest. This method allowed us to derive a ranking of the importance of all regulator expressions for the target by averaging the scores over all the trees of the random forest. The randomized subset of regulators allowed us to avoid the local minima of the global score, and the random subsample of the data used for each tree avoided over-fitting of the data and hence permitted more robust estimates. We tested on simulated data whether including auto-loops in the model improved the performance. Results are presented in Supporting Method and they show that no gain was obtained with such modified version of our RF algorithm. Compared to previously developed tree ensemble methods, our method is novel because our modeling explicitly accounted for the dynamical and multi-condition aspects of the data.

The second method used a Gaussian graphical modeling (GGM) approach. In the GGM paradigm, an edge was inferred when a significant partial correlation was detected between the expressions profiles of two genes. Namely, the partial correlation between two genes is the correlation between the residuals of the expressions of these two genes after accounting for all other gene expressions patterns. A unique aspect of our approach is the combination of a temporal approach with a multiple graph structure inference scheme. The dynamic nature of the data allowed us to obtain directed edges between two genes (i.e., changes in the expression of gene *p* induced changes in the expression of gene *q* and not the converse). In addition, the multiple graph framework drove the inference of condition-specific networks. However, each of these hormonal networks took into account information from the others and therefore accounted for a coupled functioning of the biological mechanisms that they encoded. The details of the RF and GGM approaches are provided in the Supporting Methods. For each of the ten GRNs, we selected only edges confirmed by both methods. The union of the nine hormonal consensus networks and the global consensus network formed a final unified network with hormone-specific edges and global edges.

Topological parameters

The topology of a GRN depicts the relative positions of the genes in the network and their importance in the structure of the network. The topological parameters for each node therefore represent quantitative measures of gene connectivity and network position; these parameters are calculated from the oriented edges that connect one gene with another. The edge count, the indegree and the outdegree are three correlated parameters indicating the total number of edges (in and out) and the number of outgoing and ingoing edges respectively. The average shortest path length of a node *p* is the average length of the shortest path between *p* and any other node. The closeness centrality is the reciprocal of the average shortest path length. The eccentricity is the maximum non-infinite length of the shortest path between *p* and another node in the network. As the network is directed, if *p* is a node without outgoing edges, the values of the average shortest path length, the closeness centrality, and the eccentricity could not be calculated. The betweeness centrality of a node *p* is the number of shortest path from a node *q* to a node *r* (differents from *p*) divided by the number of shortest paths from *q* to *r* that pass through *p*. It reflects the amount of control that the node *p* exerts over the interactions of other nodes in the network. The stress centrality of a node *p* is the number of shortest paths passing through *p*. Finally, the neighborhood connectivity of a node *p* is the average connectivity of all neighbors of *p*. These different metrics were calculated for all genes with the NetworkAnalyzer plugin for Cytoscape (Assenov *et al.*, 2008).

Correlation between topological parameters and genetic differentiation

First, we performed firstly a principal component analysis (PCA) on the topological parameters of the GRN to study the dependency of those parameters, with the function *princomp*. This allowed us to identify the components explaining the most parameters variability. From these PCA results, we selected the most representative topological parameters in order to avoid redundancy. The $F_{ST}$ values were grouped into 5 subsets, each of them expressing the $F_{ST}$ between one *Helianthus* population (Wild *H.annuus*, Landraces, Elite, *H.argophyllus*, *H.petiolaris*) and the other populations. We performed a canonical correlation analysis (R function *cancor*) in order to identify the canonical correlations between the selected topological parameters on one side and each $F_{ST}$ subset on the other side. We tested their significance with the test of Wilks as provided by the function *p.perm* of the R package CCP with 10 000 permutations.

## Results:

### Gene selection to infer the drought GRN

*Gene identification using a global transcriptomic approach*

To identify genes that play a role in the drought GRN, a global transcriptomic approach was employed using an Affymetrix chip containing 32,423 probesets, which corresponded to sequences expressed in *H. annuus*. The differential analysis identified 337 genes that responded to drought stress under field conditions and 447 genes that responded to drought stress under greenhouse conditions (Rengel *et al.*, 2012). Because ABA is the major plant hormone involved in the drought-stress response, we also identified genes displaying differential expression 6 hours after ABA treatment at the plantlet stage under hydroponic conditions, using a similar global transcriptomic analysis. A total of 463 sunflower transcripts were found to be differentially expressed after ABA application (Supporting Information Table S1). The 463 ABA-regulated sunflower genes were validated by comparison with the expression of 226 homologues in *Arabidopsis* based on expression data from the Bio-Array Resource database or in projects from the AtGenExpress Consortium retrieved on the website http://www.weigelworld.org/resources/microarray/AtGenExpress/AtGe_Abiostress_gcRMA.zip. The authors employed a kinetic analysis of three time points to assess the transcriptomic response to abiotic stresses such as cold, osmotic, salt, drought or heat stress in leaves using the *Arabidopsis* Affymetrix ATH1 microarray. This study was of particular interest because its kinetic approach imparts greater statistical power and avoids the issue of differences in kinetic parameters between sunflower and *Arabidopsis*. The *Arabidopsis* homologs of the sunflower genes in this study are all BLAST reciprocal best hits between *Helianthus* ESTs and *Arabidopsis*. The covariance analysis (ANCOVA) showed that the expression modulation by abiotic stresses over time of 27% of these *Arabidopsis* homologues (60 genes) exhibited a treatment effect or a treatment x time interaction effect. This proportion of Arabidopsis genes homologous to Helianthus genes responding to ABA corresponds to a significant enrichment in Arabidopsis genes responding to abiotic stresses (hypergeometric test giving $p=1.10^{-4}$). The ANCOVA analysis, hypergeometric test and results are described in detail in the Supporting Methods and Supporting Information Table S2, respectively. This finding confirms that at the transcriptomic level, ABA regulation and its role in abiotic stress responses are globally conserved between *Arabidopsis* and *H.annuus*, as it has been documented in many plants; this conservation has occurred even though sunflowers are a very distantly related lineage separated by more than 90 million years of evolution (Chinnusamy *et al.*, 2004).

These three lists contain gene groups that respond to two drought stress intensities and ABA application (mimicking a third drought stress condition) at different developmental stages. Together, they provide complementary views of the drought-regulated genes in sunflower.

For inclusion in the GRN for drought stress, we stipulated that the genes must respond to at least two of the following conditions: (1) drought stress under field conditions, (2) drought stress under controlled greenhouse conditions, and/or (3) ABA under hydroponic conditions (Fig. 1). As expected from the large variability of the biological material used to select the genes, the selected intersection was robust and should comprise the genes composing the core GRN for drought stress.

In addition to these groups of genes, we selected 56 genes that are known from the literature or gene ontology (GO) analysis to be regulated in response to ABA or one of the other main plant hormones used for the treatment in our hydroponic experiment.

In all, 181 genes were selected (see complete list of sunflower transcripts, *Arabidopsis* homologs and annotations in Supporting Information Table S3).

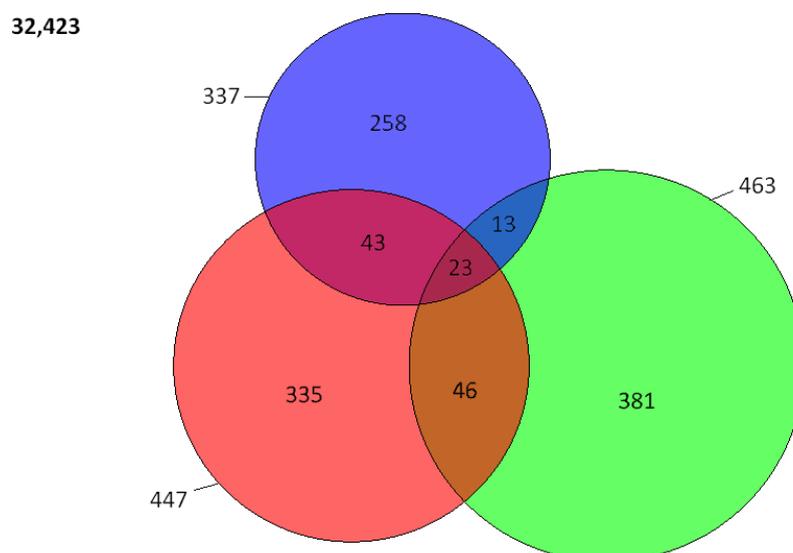

Figure 1: Selection of genes likely to be involved in the drought GRN. Genes that responded to drought stress under field conditions, drought stress under greenhouse conditions, and ABA application under hydroponic conditions are indicated in blue, red, and green, respectively. The genes that were responsive under at least two of the different conditions were selected as part of the inferred GRN for drought-stress responses.

*Dissection of transcriptional regulation in the drought GRN by application of hormonal treatments*

The GRN of these drought-regulated genes was reconstructed from their expression levels measured by q-RT-PCR. To perturb the network and identify regulatory relationships, leaf samples were harvested at seven different times after hormone treatment from hydroponically grown plants. A total of nine different hormones representing the main plant hormone groups were used. From the 181 selected candidate genes, we retained 145 robust genes based on technical filtering (efficiency, imputable missing data). The expression levels (expressed as ΔΔCt in reference to 5 control genes and the mock control) before and after imputation of missing data are shown in Supporting Information Table S4 and Supporting Information Table S5, respectively.

Inference of the drought GRN from the GGM and RF methods

*Inferences of a global GRN and nine hormonal GRNs lead to the identification of a robust unified drought GRN*

To identify the final regulatory network between the 145 genes shown to be co-expressed during drought stress, we studied their regulation after several hormonal applications. This strategy was chosen because the environmental signal is transduced by different hormones whose regulatory pathways are very connected. The application of different hormones can reveal hormone-specific and global regulatory connections. Because we selected genes shown to respond to drought, the revealed regulatory connections are likely involved in drought-stress responses. We generated nine datasets corresponding to the nine hormonal treatments and containing expression values for the 145 robust genes at seven different time points. From these nine datasets, we established 10 GRNs: one GRN from each hormonal treatment and one global GRN, which represents a consensus array of all hormonal treatments. The GRNs were inferred using two different inference methods: Gaussian graphical modeling (GGM) and random forest (RF). These two approaches produce complementary predictions (Allouche *et al.*, 2013), and merging their results was shown to yield more reliable predictions than predictions obtained by any single method (Marbach *et al.*, 2012).

With the GGM method, we obtained between 112 and 158 edges for each hormonal network and a global network with 95 edges (Fig. 2).

With the RF method, the number of edges for each hormonal network was very different and varied from 11 to 174 edges. The global GRN with the RF inference was composed of 242 edges (Fig. 2).

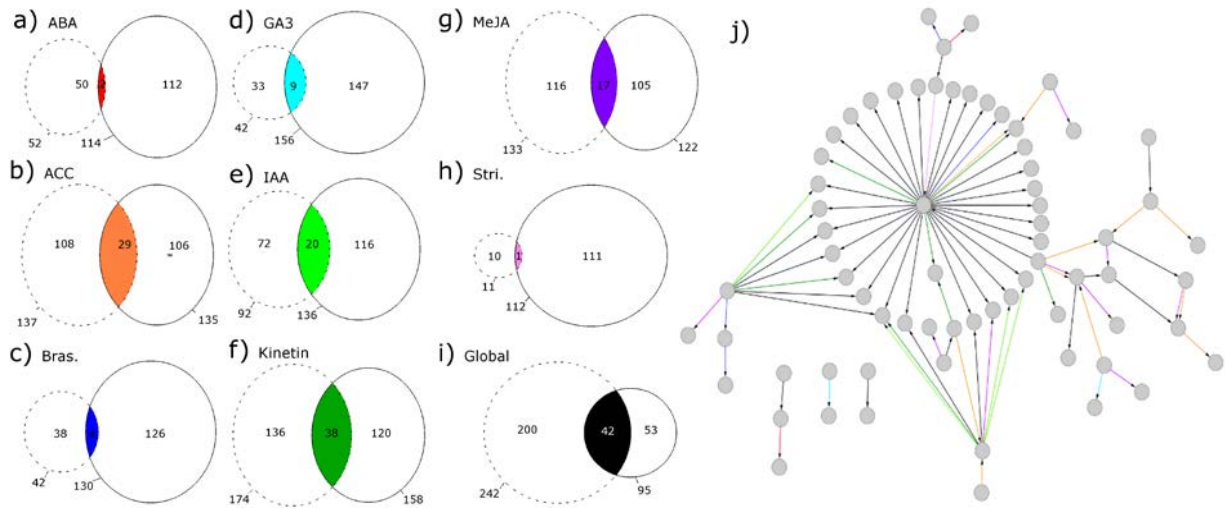

Figure 2: Drought GRN and selection of its edges. a-i: The Venn diagrams for each hormonal GRN and global GRN represent the edges selected by the RF method (dotted line) and the GGM method (solid line). a) ABA. b) Ethylene. c) Brassinosteroid. d) Gibberellin. e) IAA. f) Kinetin. g) Methyl-jasmonate. h) Strigolactone. i) Global. j) Unified drought GRN representation. Grey circles represent the genes. Arrows represent the relationships between two genes (oriented edges), and their color represents the hormonal treatment that led to their identification: Red = ABA; Orange = Ethylene; Dark blue = Brassinosteroid; Light blue = Gibberellin; Light green = IAA; Dark green = Kinetin; Violet = Methyl-jasmonate; Pink = Strigolactone; and Black = Global or non-hormone-specific edges.

Given the diversity in the inferred edges, we employed a very stringent approach to retain the core, most robust GRN. First, we discarded the results of SA treatment because the RF method inferred 629 edges. This number was far higher than that for the other hormones (49, 115, 38, 36, 94, 134, 147, and 16 when including SA). We chose not to take into account the SA edges in the final GRN to avoid an over-representation (more than 25%) of specific edges for this hormone instead of drought edges. Second, for each GRN (hormonal or global), we considered an edge to be robust if it was selected by both the GGM and RF methods. This is a conservative approach that leads to high-quality edges; we chose to focus on a network with very reliable edges at the expense of potentially missing some weaker associations that might be relevant. This trade-off was confirmed in very different scenarios based on both simulated and real data sets (Vignes *et al.*, 2011; Marbach *et al.*, 2012). We validated both our models using simulated data that had the specific features of the data being studied (see the Supporting Methods). Note that the numbers of robust edges were very different depending on the focal GRN. The final unified network, hereafter called the drought GRN, was formed by the union of all these robust edges (Fig. 2) and comprised 69 connected nodes, representing the genes linked by 79 unique edges. Among the 69 genes, 49 were differentially expressed in one of the three global transcriptomic experiments using the Helianthus Affymetrix chip, and only 20 came from the literature or GO analyses using BLAST reciprocal best hits to infer homology. Supporting Information Figure S1 summarizes the origins of the 69 final genes of the network.

The number of shared edges between the hormonal GRNs varied from 0 to 18 (Supporting Information Fig. S2 and Table S6). The ethylene, cytokinin, and auxin networks shared the largest number of edges, whereas the ABA, brassinosteroid, and strigolactone networks had no edges in common with the other hormonal networks.

*Comparison of the drought GRN to* Arabidopsis *data and prior knowledge of biological networks*

We compared our sunflower drought GRN to the model plant *Arabidopsis thaliana* using expression data from the AtGenExpress Consortium (Goda *et al.*, 2008) (GEO accession: GSE39384 from AtGenExpress Consortium). This *Arabidopsis* data set was similar to the *Helianthus* data and includes seven hormonal treatments but is limited to only three time points. Due to this difference in the sampling frequency, we were unable to define a network from these data using the inference methods described above. Therefore, we searched for gene expression correlations that were consistent (or inconsistent) with the sunflower data. Among the 116 *Arabidopsis* genes that were homologous to the 145 sunflower genes that were initially used to develop the consensus drought GRN, significant correlations between gene pairs were more frequent for pairs corresponding to the network edges, according to an exact hypergeometric test (p=0.005). The correlation analysis and hypergeometric test are described in the Supporting Methods. This result demonstrated that the gene expression correlations identified from the *Arabidopsis* data were similar to the correlations identified in our sunflower drought GRN.

The topology of the drought GRN is consistent with what is known about biological networks. The degree distribution of the sunflower drought GRN followed a power law $y = 20.57x^{-1.98}$ with an $R^2$ of 0.72 (Supporting Information Fig. S3). This means that a few nodes had many connections and that the majority of the nodes had few edges, a finding that is a typical feature of the scale-free topology of biological networks (Barabasi & Oltvai, 2004).

## Node connectivity defines different gene classes

### *Identification of two hubs sharing common targets*

The average value for the connectivity of a node (i.e., the number of outgoing or ingoing edges connecting a node to the others) in the inferred drought GRN was 2.3. However, we identified nodes with important connectivity; in particular, two nodes had the highest number of outgoing edges: 8 and 32 (with a connectivity of 9 and 32 respectively). These two genes were identified as important hubs in the inferred GRN. In addition, these genes shared 7 common targets, while no common sources (i.e., a gene $q$ that targets the studied gene $p$) between these genes were identified.

### *Relation between connectivity and gene function*

Gene ontology annotations of the *Arabidopsis* genes homologous to the 69 *Helianthus* genes connected in the unified drought GRN were retrieved from TAIR based on protein homology using the sunflower transcriptome web portal (www.heliagene.org/HaT13l). We observed that genes in the GO metabolism category accounted for the majority of the genes with low connectivity values: 40%, 80% and 60% of the genes with a connectivity of one, two and three respectively, however there was no significant enrichment using a hypergeometric test ($p=0.190$). More interestingly, genes annotated as transcription factors and as having DNA-binding properties exhibited medium connectivity (i.e., four to five edges, $p=0.002$), with the exception of one gene that had a single edge, possibly because its targets were filtered out during our analysis. Finally, the most highly connected genes were anion transporters. While the GO transporter included 20-30% of the genes with low connectivity, it also contained all the genes with high connectivity, including both hubs, which had 9 and 32 edges (Fig. 3). The test showed that despite the very low number of highly connected genes, this trend was significant ($p=0.059$).

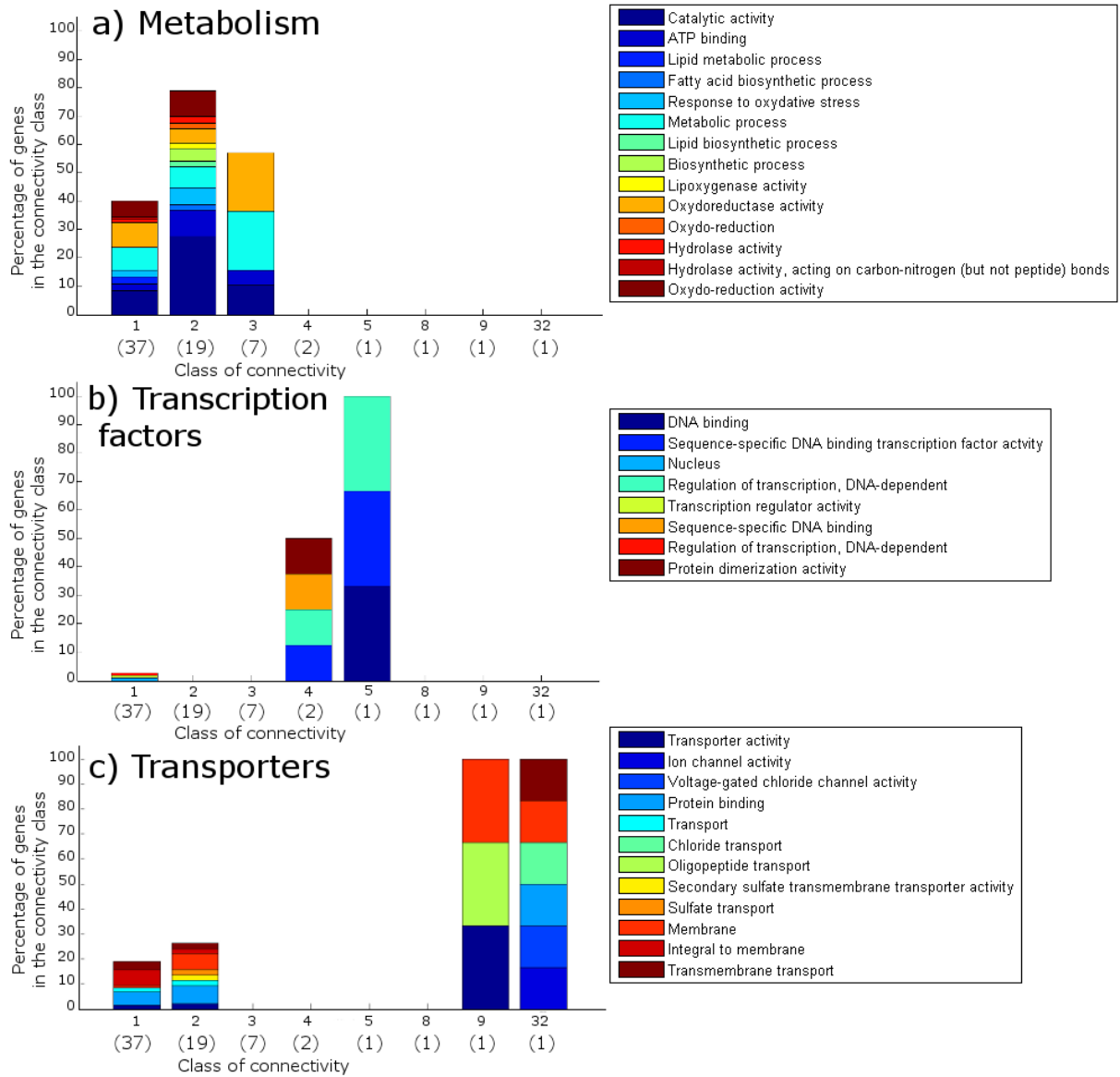

Figure 3: Percentage of GO terms for each connectivity class of genes in the drought GRN, where the GO are represented by different colors. a) Metabolism. b) Transcription factor or DNA binding. c) Transporters. Number of genes in each connectivity class is indicated between brackets. Note that the connectivity classes of 8 is represented by a unique gene which does not belong to any of the three main classes of GO represented here (metabolism, transcription factor and transporters).

Canonical correlations between the topological parameters of the drought GRN and genetic differentiation statistics

To examine how the drought GRN might be related to the evolution of wild and domesticated sunflower populations, we looked for canonical correlations between non redundant network topology parameters and the genetic differentiation statistics of the drought GRN nodes or genes. The topological parameters for each node represent quantitative measures of the gene position and relationships to others in the network. They are calculated from the number of oriented edges that connect one gene with another and are not independent by construction. In our GRN, edges are oriented, thus, we only considered genes with outgoing edges to compare the predictive value of the topological parameters. In addition, we were able to calculate $F_{ST}$ for 15 of these genes among five populations of *Helianthus*: wild *H. annuus*, landrace lines of *H. annuus*, elite lines of *H. annuus*, *H. petiolaris*, and *H. argophyllus*.

|  | Rho Correlation coefficient 1 | Rho Correlation coefficient 2 |
|---|---|---|
| $F_{ST}$ subset of *H. argophyllus* | 0.672 (p-value= 0.299) | 0.524 (p-value: NS) |
| $F_{ST}$ subset of *H. petiolaris* | 0.493 (p-value=0.818) | 0.369 (p-value: NS) |
| $F_{ST}$ subset of *H. annuus* Wild | 0.728 (p-value= 0.362) | 0.292 (p-value: NS) |
| $F_{ST}$ subset of *H. annuus* Landraces | 0.976 (p-value= $1 \times 10^{-4}$) | 0.299 (p-value: NS) |
| $F_{ST}$ subset of *H. annuus* Elite lines | 0.946 (p-value=0.002) | 0.280 (p-value: NS) |

Table 1: Coefficients of canonical correlations between in one hand, topological parameters values of the drought GRN nodes and in another hand, their genetic differentiation measured as $F_{ST}$ and grouped in five subsets. Each subset of $F_{ST}$ compares genetic differentiation of one *Helianthus* population to the four other populations of *Helianthus*. Correlations were tested for significance with Wilks's test with the function *p.perm* of the R software computing 10 000 permutations. NS= non significant.

In a first step we used results from the PCA (cf Supporting Figure S4 and Table S7) with topological parameters to reduce dimensionality and to obtain independent variables. The first and second components explained 67% of the variance. Regarding their loadings on the first two principal components, we selected ASPL and EdgeCount (cf Supporting Table S7). Genetic differentiation was analyzed using five distinct $F_{ST}$ subsets each of them expressing the $F_{ST}$ between one *Helianthus* population and the other populations. Canonical correlation analysis (Table 1 and Supporting Table S8) between each of these five $F_{ST}$ subsets on one side, and the two topological variables selected on the other side allowed to detect significant canonical correlations only for the Elite $F_{ST}$ subset (Wilks's test p= $2.00 \times 10^{-3}$) and for the Landrace $F_{ST}$ subset (p= $1.00 \times 10^{-4}$). As the intersection between these two subsets was $F_{ST}$ between Elite and Landrace, this suggests that this variable in particular is correlated to the topological properties of the GRN. It was confirmed by the comparison of the canonical correlation analyses including only the $F_{ST}$ value between Landraces and Elite lines (Wilks's test p =$1.90 \times 10^{-3}$) or the $F_{ST}$ value between Landraces and Wild (Wilks's test p = 0.26). More specifically, we found a significant correlation of Pearson between $F_{ST}$ value between Landraces and Elite lines and ASPL (R = 0.74, p=0.003).

**Discussion:**

In this study, we reconstructed a GRN based on gene expression that portrays the transcriptional regulations that occur within a plant organ in response to environmental cues. As such, this drought GRN is not based on physical interactions between gene products and promoters and thus is not a molecular cell biology model. Instead, this GRN provides a more physiological view based on transcriptional events involved in drought stress responses similarly to the study of (Hannah *et al.*, 2006) on freezing tolerance in Arabidopsis. In addition, due to the temporal approach, the network edges are oriented and can be interpreted as dependent relationships. Together, these characteristics produce a network based on molecular regulations that also integrates physiological processes with their chronology at the organ level. This provides a representation of plant physiological responses to dry conditions and therefore of the fitness in such an environment.

Network inference highlights the importance of nitrate transport in guard cells

*Drought GRN hubs are nitrate transporters and drive transcriptional regulation*

In the inferred network, two genes had many outgoing connections compared with other genes and could therefore be considered hubs. The first hub (HaT13l030730) is homologous to the transcript of the *Arabidopsis* gene chloride channel A (CLC-A, AT5G40890). CLC family members are involved in anion compartmentalization in intracellular organelles and in stomatal guard cell vacuoles (Jossier *et al.*, 2010). More precisely, CLC-A and CLC-C are expressed in stomata and control their opening through translocation of $NO_3^-$ and $Cl^-$, respectively. This difference in anion selectivity among the CLC family members is due to an amino acid change in the selectivity filter (Wege *et al.*, 2010). The sunflower transcript HaT13l030730, which is homologous to *Arabidopsis* CLC-A, possesses the same amino acid conferring nitrate specificity. This suggests that the main hub identified in the drought GRN is likely a nitrate channel involved in stomatal aperture control and, therefore, transpiration.

The second hub (HaT13l003541) is homologous to the transcript of the *Arabidopsis* gene NRT1.1 (AT1G12110), which encodes a dual-affinity nitrate transporter in *Arabidopsis*. Guo et al. (Guo *et al.*, 2003) demonstrated that this gene is expressed in guard cells of stomata and that transpiration is affected in mutants in an ABA-independent manner. The reduction of the stomatal aperture in mutants appeared to be due to nitrate uptake in guard cells. The control of stomatal transpiration by anion channels and transporters in guard cells was further confirmed (De Angeli *et al.*, 2013) in *Arabidopsis*.

Our approach identified the key role of two sunflower homologues of *Arabidopsis* anion transporters. This strongly suggests that this process is important for the regulation of the sunflower drought response. However, the two hubs do not directly regulate the expression of their target as transcription factors do; instead, the hubs drive downstream signaling cascades through indirect physiological and distant regulations.

*The drought GRN identifies connections between ABA-dependent and ABA-independent pathways*

In the inferred network, both hubs had seven common targets but no common source. This suggests that the NRT1.1 and CLC-A sunflower homologues could represent two pathways controlling drought stress responses. However, we could not exclude a cross-talk between NRT1.1 and CLC-A with an upstream regulator absent from our initial dataset. By inferring sunflower gene function based on *Arabidopsis* homology and the analogous expression response to drought, we could tentatively investigate the molecular pathways characterized in the sunflower drought GRN. Functional annotation of the targets of the two hubs revealed genes that are directly involved in cell protection and stress tolerance, such as the ROS scavenger (APX1) and two enzymes involved in synthesis of an osmo-protectant, choline (PMEAMT and CCT2). Interestingly, we also identified genes involved in signal transduction, such as kinases (HaT13l074901 and emb1075), phosphatases (HAB1), calmodulin-binding proteins (CPK5), and transcriptional regulators (MYC2, ARIA), downstream of the anion transporters, as described in Fig. 4.

CLC-A and NRT1.1 define an ABA-dependent and an ABA-independent, respectively, pathway in our experimental results, as well as in *Arabidopsis* (Guo *et al.*, 2003; Jossier *et al.*, 2010). Both sources of CLC-A, SULTR1 and ABF2, are regulated by ABA in *Arabidopsis* (Fujita *et al.*, 2005), (Ernst *et al.*, 2010) and also in our experiment for ABF2. In addition, specific targets of CLC-A are part of the ABA signaling cascade in *Arabidopsis*. HAB1 is a protein phosphatase that is strongly up-regulated by ABA (Rodriguez, 1998) and functions in ABA signaling. ABA1 is known to catalyze the first step of ABA synthesis (Rock & Zeevaart, 1991), and ARIA is an armadillo repeat protein that is known to interact with the transcription factor ABF2 (Kim *et al.*, 2004). Together, these regulatory connections identified in *Arabidopsis* form a loop involving ABA synthesis (in vascular cells) (Boursiac *et al.*, 2013) and a signaling pathway across the different cell types (including guard cells) throughout the leaf (Fig. 4). In the drought GRN, we were able to partially identify the corresponding regulatory loop between sunflower homologues. These results suggest that the same ABA regulatory loop exists in the sunflower drought GRN and therefore could be largely shared across the plant kingdom.

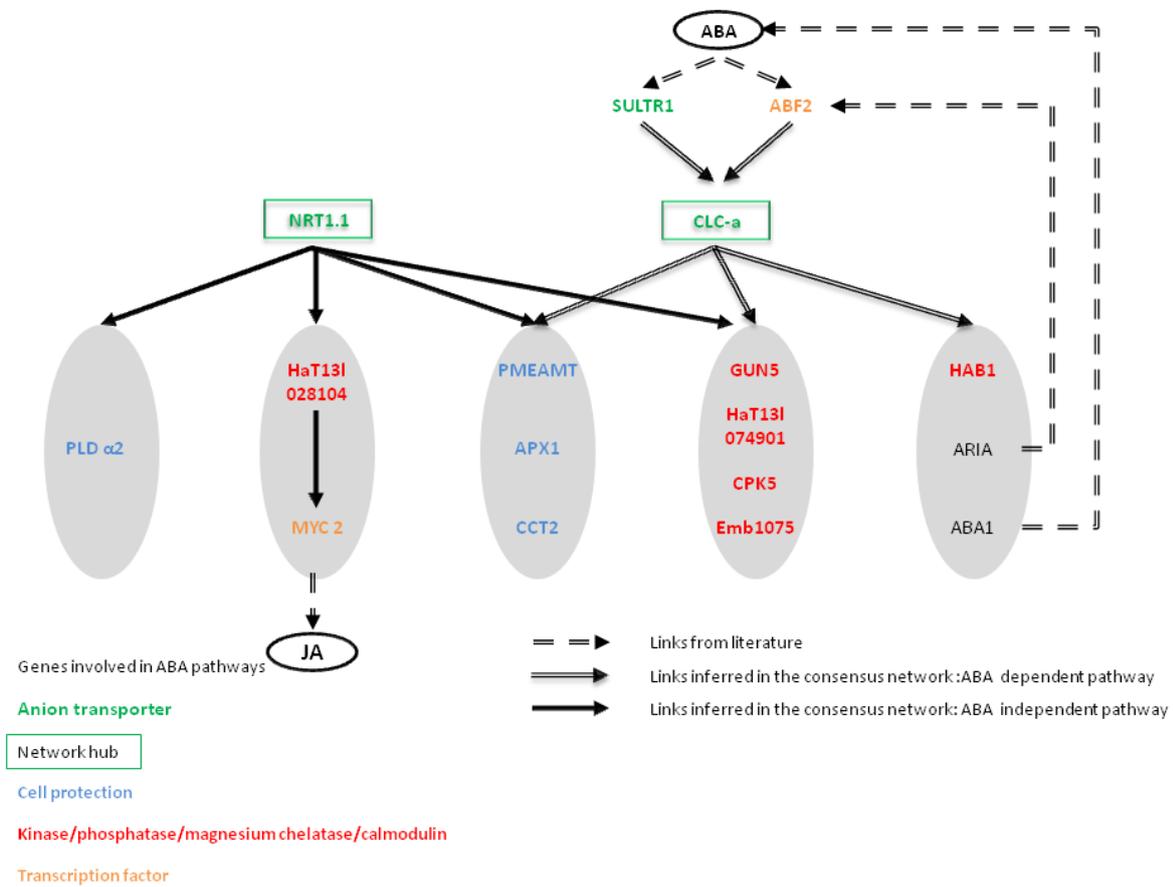

Figure 4: Functional network involving the two hubs of the inferred drought GRN, their sources, and their targets. Blank edges represent the ABA-dependent pathway, including the CLC-A. Solid edges represent the ABA-independent pathway, including NRT1.1. Common targets involved in signal transduction are indicated in red, those involved in transcriptional regulation are shown in orange, and those involved in cell protection are shown in blue. Abbreviations in the figure: PLDα2: Phospholipase D alpha2; MYC2: MYC related transcriptional activator; PMEAMT: phosphoethanolamine n-methyltransferase; APX1: ascorbate peroxidase 1; CCT2: phosphorylcholine cytidylyltransferase2; GUN5: genomes uncoupled 5; CPK5: calmodulin-domain protein kinase 5; emb1075: embryo defective 1075; HAB1: hypersensitive to aba1; ARIA: arm repeat protein interacting with abf2; ABA1: aba deficient 1; CLC-a: chloride channel a; NRT1.1: nitrate transporter 1; SULTR1: sulfate transporter 1; ABA: acid abscisic; JA: jasmonate.

Similar to the shared targets of CLC-A and NRT1.1, specific targets of NRT1.1 are also involved in cell protection (PLDα2) and signal transduction (HaT13l028104). An interesting downstream target is MYC2, which is a central regulator of the hormone jasmonate, which is mostly involved in plant defense and the development and integration of many hormonal signals (Kazan & Manners, 2013).

Across the sunflower drought GRN, several different pathways show some conservation across plant species, such as *Arabidopsis*. Therefore, the GRN inference approach developed in this study appears to be robust, and we can make the reasonable hypothesis that the main regulatory pathways and hubs identified in the drought GRN are likely conserved among distant plant species and therefore also across the *Helianthus* genus. Although, from our data we were not able to demonstrate the network conservation across *Helianthus* population (it would require inferring the network for each one which would be too laborious with the present technologies), this hypothesis allows us to explore new questions about how the GRN could constrain plant adaptation to dry environments.

Drought GRN topology and *Helianthus* evolution
*Network topology constrains genetic variation of the gene network*
Gene networks are the products of evolution, similarly to other biological objects, but gene network relationships can also constrain evolutionary changes, such as adaptations to new environments and responses to selective pressure during domestication or breeding. For example, Rausher *et al.* (1999) demonstrated different evolutionary histories for upstream and downstream genes in the anthocyanin biosynthetic pathway.

The evolution of the GRN architecture can lead to new nodes, potentially introducing new functions and new edges between these nodes. Previous researchers (Hinman *et al.*, 2003) examined GRN evolution in echinoderms and demonstrated that some features of developmental GRNs were conserved and that others were specific to each taxa. Network architecture is known to affect evolutionary rates (Ramsay *et al.*, 2009), and we expect evolutionary changes to the nodes to be constrained by their connectivity and the number of neighbors. A hub in the network is involved in several pathways. The functional trade-offs for such genes are higher than those for peripheral genes that are neither involved in regulatory processes nor in the interaction with partners.

To understand how populations and species evolve and adapt to a new environment, we examined the putative constraints of the network architecture on the genetic differentiation between populations of *H. annuus,* and two wild species that are cross-compatible with *H. annuus*: *H. argophyllus* and *H. petiolaris*.

*No evidence of network topology constraints during the divergence of H. argophyllus and H. petiolaris*

*Helianthus argophyllus* is native to the dry, sandy soils of southern Texas, an arid environment that imposes strong selection for tolerance to drought stress. Indeed, *H. argophyllus* is considered the most drought-tolerant sunflower species because its pubescent leaves reflect sunlight, reduce water loss, and exhibit low transpiration (Seiler & Rieseberg, 1997). However, network topology and $F_{ST}$ values between *H. argophyllus* and other populations were not significantly correlated. This could be because the adaptation of *H. argophyllus* to dry environments involved physiological mechanisms that are not captured in our GRN or because the network topology has itself evolved and the topological parameters in *H. argophyllus* are too dissimilar to those in *H. annuus*. Interestingly, the highest value of $F_{ST}$ between *H. argophyllus* and other populations was for the network hub, NRT1.1, which is involved in transpiration. This result is consistent with positive selection acting on NRT1.1 during adaptation of the *H. argophyllus* to dry environments. Keeping in mind the overall non-significant correlation, it suggests that NRT1.1 could be an example of the fore-mentioned hypothesis.

In *H. petiolaris,* we observed no correlation between the GRN topology and $F_{ST}$ for comparisons with other populations. Because *H. petiolaris* has a large geographic range that overlaps with that of *H. annuus* in the Great Plains of the USA, drought stress might not be the major selective force separating these species. This could explain the similar divergence patterns within the drought network genes between these two populations as illustrated in Fig5B.

*Genetic diversity within the GRN was modified during modern breeding*

The network topological parameters and the $F_{ST}$ between the landraces and elite lines of Helianthus annuus were correlated (Fig5A). This reflects a difference of genetic differentiation between these two populations between the center and the periphery of the network. We did not observe this correlation for $F_{ST}$ between wild H. annuus and landraces. This suggests that the position and connectivity of genes in the drought GRN influenced the response to selection during the last century of genetic improvement but not during the initial domestication of H. annuus. This difference in selective responses could be due to the fact that highly connected genes are subjected to more trade-offs as they are master regulators with involvement in several genetic pathways in contrast to less connected terminal genes (Fig5A). Drought tolerance is considered to be a long standing goal of sunflower breeders. We would expect that the selection they exert had led to a global reduction of genetic diversity in the drought GRN. However, we observed a higher divergence of terminal genes compared to central ones, which implies a stabilizing selection acting on the network hubs. Interestingly, our $F_{ST}$ studies in H. argophyllus, suggest that a different selective pressure acted on one of the network hub (Fig5B). This highlights our lack of global understanding on how evolutionary forces and functional relationships interacted to produce contemporary phenotypic diversity and suggests a potentially important way of improving the breeders' methods, through the integration of regulatory networks in quantitative genetics models such as genomic selection.

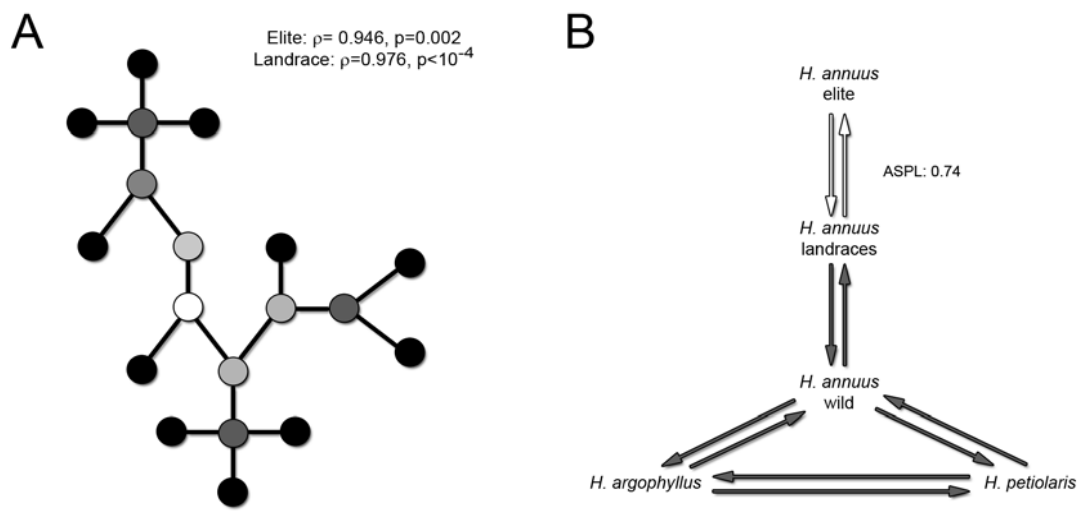

Figure 5: A. Representation of genetic differentiation between *H. annuus* Landraces and *H. annuus* Elite Lines in function of the gene positions in a schematic gene regulatory network. Colors of the genes in the schematic GRN represent the difference of heterozygocity between the two populations for the considered gene. Node color represents differentiation between elite and landrace sunflowers: darker nodes appeared more differentiated compared to lighter nodes. Canonical coefficients (ρ) and p-value of the Wilks's test for the correlations between network topological parameters and $F_{ST}$ values of one population compared to the others are indicated for Elite lines and landraces. B. Hypothesis about differences of genetic differentiation between the five *Helianthus* populations. Note that only the five comparisons representing the selective history of the sunflower are shown. Black edges indicate no variability in the genetic differentiation within genes network between the two populations. White edges indicate changes in the genetic differentiation between populations as observed for the 15 genes in the drought GRN analyzed in the CCA. The coefficient of the Person's correlation between the topological parameter Average Shortest Path Length (ASPL) and $F_{ST}$ between *H. annuus* Elite lines and Landraces is indicated.

In conclusion, this work investigates the interaction between physiological and evolutionary processes in the context of a genetic network for the drought-stress response. Interactions between physiological and evolutionary time scales could be revealed in the future through global transcriptomic studies, although some limitations of network inference methods remain to be overcome. This type of work will facilitate the study of responses to other environmental factors and clarify whether physiological mechanisms and evolutionary adaptation, which are reciprocally constrained in the gene regulatory network, are similar in abiotic and biotic interactions.

**Acknowledgments:**

This work was part of the OLEOSOL project funded by French public funds for competitiveness clusters (FUI), the European Regional Development Fund (ERDF), and the Government of the Région Midi-Pyrénées. The study benefited from the support of the Genoscope project AP09/10 and a PhD grant co-funded by Syngenta Seeds and the Région Midi-Pyrénées. This work was part of the "Laboratoire d'Excellence (LABEX) entitled TULIP (ANR -10-LABX-41)".

We thank the Platform GeT PlaGe from INRA Toulouse for their technical support and the sunflower team members and the common services of the LIPM.

**Supporting Information:**

**Methods S1:** Supporting methods providing detailed information about plant material, molecular biology procedures and GRN inference.

**Table S1:** Results of t-test demonstrating the differential expression of genes upon application of 10 µM ABA.
**Table S2:** Results of ANCOVA showing the validation of the ABA genes dataset.
**Table S3:** Description of the genes selected for GRN inference.
**Table S4:** Raw gene expressions.
**Table S5:** Gene expressions after log transformation and missing data imputation.
**Table S6:** Number of edges detected for each hormone.
**Table S7:** Complete results of the Principal Component Analysis on the topological parameters for the drought GRN.
**Table S8:** Complete results of the Canonical Correlation Analysis between the topological parameters and $F_{ST}$ subsets.

**Figure S1:** Origin of the selection for the inferred genes of the drought GRN.
**Figure S2:** Hormonal and Global Networks.
**Figure S3:** Degree distribution.
**Figure S4:** Bi-plot of the effects of the topological parameters in a Principal Component Analysis. Components 1 and 2 are shown.